\documentstyle[11pt,newpasp,twoside,psfig]{article}
\def\edcomment#1{\iffalse\marginpar{\raggedright\sl#1\/}\else\relax\fi
}

\reversemarginpar

\begin{document}
\title{X-ray Spectroscopy of BAL and Mini-BAL QSOs}
\author{S. C. Gallagher, W. N. Brandt, G. Chartas, \& G. P. Garmire}
\affil{The Pennsylvania State University, Department of Astronomy \& Astrophysics, 525 Davey Laboratory, University Park, PA 16802}
\begin{abstract}
BAL QSOs are notoriously faint X-ray sources, presumably due to 
extreme intrinsic absorption.  However, several objects have
begun to appear through the obscuration with recent X-ray observations
by \emph{Chandra} and \emph{ASCA}.  Where enough counts are present for X-ray 
spectroscopy, the signatures of absorption are clear.  The evidence
is also mounting that the absorbers are more complicated than
previous simple models assumed; current absorber models need to be extended 
to the high-luminosity, high-velocity, and high-ionization regime 
appropriate for BAL QSOs.
\end{abstract}
\section{Introduction}
Since the first surveys with \emph{ROSAT}, BAL~QSOs have been known to be
faint soft X-ray sources compared to their optical fluxes (Kopko,
Turnshek, \& Espey 1994; Green \& Mathur 1996).  Given the extreme
absorption evident in the ultraviolet, this soft X-ray faintness was
assumed to result from intrinsic absorption.  Based on this model, the 
intrinsic column densities required to suppress the X-ray flux, assuming a
normal QSO spectral energy distribution, were found to be
$\ga5\times10^{22}$~cm$^{-2}$ (Green \& Mathur 1996).  Due to the 2--10~keV
response of its detectors, a subsequent \emph{ASCA}
survey was able to raise this lower limit by an order of magnitude for
some objects, to $\ga5\times10^{23}$~cm$^{-2}$ (Gallagher et al. 1999).  
In all of these studies, the premise of an underlying typical QSO
spectral energy distribution and X-ray continuum was maintained.  The
strong correlation found by Brandt, Laor, \& Wills (2000) between
C\,{\sc iv} absorption equivalent width (EW) and faintness in soft X-rays further 
supported this assumption.
 
\emph{ASCA} observations of individual BAL~QSOs such as PHL~5200 (Mathur, Elvis,
\& Singh 1995) and Mrk~231 (Iwasawa 1999; Turner 1999) provided suggestive 
evidence that intrisic absorption was in fact to blame for X-ray
faintness.  However, limited photon statistics precluded a definitive diagnosis.  The observation
of PG~2112+059 with \emph{ASCA} on 1999 Oct 30 provided the first solid evidence
for intrinsic X-ray absorption in a BAL~QSO.
%
\begin{table}
\caption{Basic Properties of BAL and Mini-BAL QSOs\tablenotemark{\rm a}}
\begin{tabular}{lcccccc}
\tableline
Name	&	
$R$	&	
$z$	&
Intrinsic $N_{\rm H}$	&
$f_{\rm cov}$\tablenotemark{\rm b}	&
$F_{2-10}$\tablenotemark{\rm c}	&
	\\
	&
	&
	&
($10^{22}$\,cm$^{-2}$)	&
	&
($10^{-14}$\,erg\,cm$^{-2}$\,s$^{-1}$)	
\\
\tableline
APM~08279+5255\tablenotemark{\rm d}& 15.2	& 3.87	& $7.0^{+2.6}_{-2.2}$
& 	&	41	\\
RX~J0911.4+0551\tablenotemark{\rm d,e} & 18.0  & 2.80 &	$19^{+28}_{-18}$
& $0.71^{+0.20}_{-0.39}$	& 5.8\\ 
PG~1115+080\tablenotemark{\rm d,e} & 15.8  & 1.72 & $3.8^{+2.5}_{-2.2}$
& $0.64^{+0.11}_{-0.16}$	& 30	\\
PG~2112+059 &	15.4\tablenotemark{\rm f} 	& 0.457	& $1.1^{+0.5}_{-0.4}$	&
& 75 \\	
\tableline\tableline
\end{tabular}
\parbox{5.8truein}{
$^{\rm a}$X-ray errors given are for $90\%$ confidence taking
all parameters except normalization to be of interest. 
$^{\rm b}$Covering fraction is only provided when partial-covering
absorption models provided a better fit to the data than simple neutral absorption.
$^{\rm c}$Flux measured in the 2--10\,keV band from the
best-fitting X-ray spectral model. 
$^{\rm d}$Gravitational lens system.  The listed optical
magnitude is for the brightest image, and the X-ray spectral
information is for all images combined.
$^{\rm e}$Mini-BAL~QSO, see \S~3.2.
$^{\rm f}$$B$ magnitude.}
\end{table}
%
\section{X-ray Spectroscopy of a BAL~QSO: PG~2112+059}
PG~2112+059 is one of the most luminous low-redshift Palomar-Green QSOs
with $M_V=-27.3$.  Ultraviolet spectroscopy with \emph{HST} clearly
revealed broad, shallow C\,{\sc iv} absorption (Jannuzi et al. 1998)
with EW of 19\,\AA\ (Brandt et al. 2000).  A 21.1\,ks \emph{ROSAT} PSPC observation
on 1991 Nov 15 detected PG~2112+059, unusual for a BAL~QSO.  A 31.9\,ks
\emph{ASCA} observation provided $\approx2000$~counts in all four detectors, 
enough for spectroscopic analysis (described in detail in Gallagher et 
al. 2001).  

To model the continuum, the data were fit above 3\,keV (2\,keV in the observed frame) with a
power law.   The resulting photon index,
$\Gamma=1.94^{+0.23}_{-0.21}$, was consistent with those of typical radio-quiet 
QSOs (e.g., Reeves \& Turner 2000).  The power law was extrapolated
back to the lowest energies in the \emph{ASCA} bandpass to investigate potential
intrinsic absorption (see Figure\,1).  The significant negative
residuals are indicative of strong absorption, and the spectrum was
subsequently fit with an intrinsic, neutral absorber with a column density,
$N_{\rm H}\approx10^{22}$\,cm$^{-2}$.  The structure in the residuals is
suggestive of complexity in the absorption, but the signal-to-noise
ratio was insufficient to investigate this fully.

This analysis of PG~2112+059 provided the first direct evidence of a
normal  X-ray continuum suffering from intrinsic absorption
in a BAL~QSO.  In addition, correcting the X-ray flux for
this absorption also demonstrated that this BAL~QSO had an underlying 
spectral energy distribution typical
of radio-quiet QSOs.
Though PG~2112+059 has a high optical flux with $B=15.4$, many of the
optically brightest BAL~QSOs have been undetected in \emph{ASCA} observations of
similar or greater exposure time.  Notably, PG~0946+301 ($B=16.0$) was barely detected 
by \emph{ASCA} with $\approx100$\,ks (Mathur et al. 2000).

The launch of \emph{Chandra} has opened a new era in X-ray observations of
BAL~QSOs.  The low background and excellent spatial resolution allow
\emph{Chandra} to probe 2--10\,keV fluxes approximately twice as faint as \emph{ASCA} in
 $\la\onequarter$ of the time.
%
\begin{figure}[th]
%
%
\centerline{\psfig{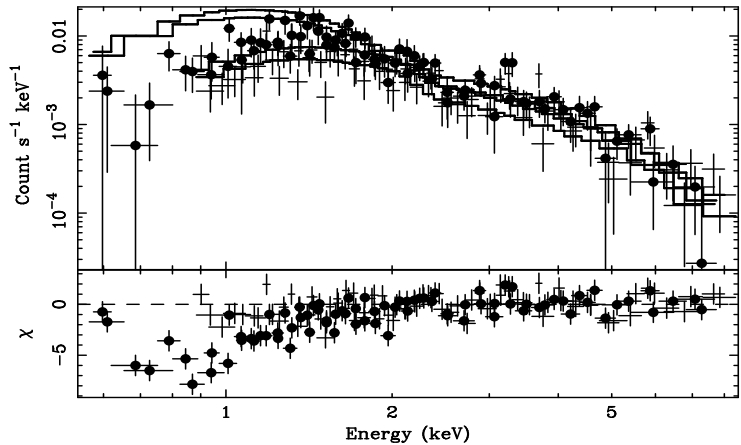}}
\caption{\emph{ASCA} SIS and GIS observed-frame spectra of
PG~2112+059 fitted with a power-law model above 2\,keV, which has then been
extrapolated back to lower energies.  Note the significant negative
residuals (lower panel) below $\approx2$\,keV suggestive of a complex absorber.}
\end{figure}
%
\section{Gravitationally Lensed QSOs with Broad Absorption}
As part of a \emph{Chandra} GTO program, gravitationally lensed QSOs
were observed with ACIS-S3 to take advantage of the power of 
the High Resolution Mirror Assembly to resolve the individual lensed images.  
As an added benefit, several of these targets contain
broad absorption lines.  The magnifying effect of the lensing allows us to
probe fainter intrinsic X-ray luminosities than would otherwise be
possible, and these targets provided
some of the best prospects for spectroscopic analysis.  The X-ray spectral
analysis for each object was done on all images combined to increase
the signal-to-noise ratio, and the results are summarized in Table~1.

\subsection{APM~08279+5255}
Since its recent discovery in 1998, the BAL~QSO APM~08279+5255 has inspired more than 20
publications.  Its apparently incredible bolometric luminosity, which seemed to 
exceed $10^{15}\,L_{\odot}$, was found to be magnified by a factor of $\ga40$ by
gravitational lensing (Irwin et al. 1998).  On 2000 Oct 11, APM~08279+5255 was
observed for 9.3\,ks 
with the ACIS-S3 intrument.  \emph{Chandra} revealed this
luminous QSO to be sufficiently bright for spectral analysis.  A
result similar to that obtained for PG~2112+059 was found: APM~08279+5255 showed a
typical QSO continuum with the signature of strong absorption.  The
best-fitting model is comprised of a power law with $7\times10^{22}$\,cm$^{-2}$
of neutral absorbing gas (see Figure~2).  Increasing the complexity of the
spectral model to include a partially covering or ionized absorber did not
improve the fits.  At such high redshift, $z=3.87$, the
diagnostics for such models have passed below the \emph{Chandra} bandpass.
%
\begin{figure}[th]
%
%
\centerline{\psfig{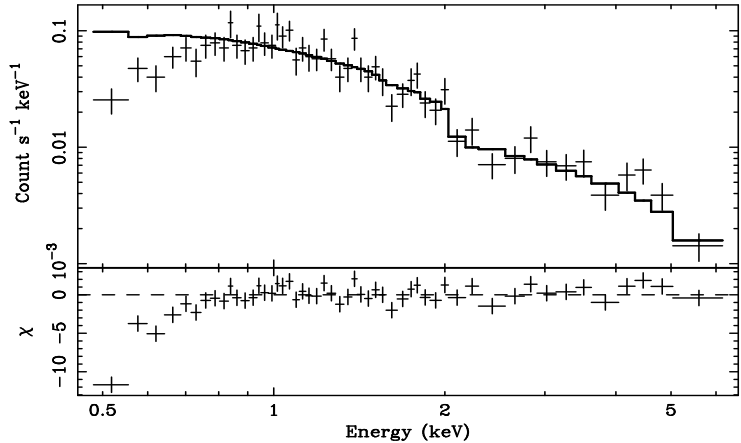}}
{\caption{\protect \emph{Chandra} ACIS-S3 spectrum of APM~08279+5255 fit with a
power law above 5\,keV (1\,keV in the observed-frame) which has then been
extrapolated back to lower energies.  The high redshift of APM~08279+5255 has
shifted the signatures of absorption almost completely out of the ACIS bandpass.}}
\end{figure}
%
\subsection{RX~J0911.4+0551 and PG~1115+080}
In contrast to APM~08279+5255, RX~J0911.4+0551 and PG~1115+080 have luminosities more
comparable to those of Seyfert~1 galaxies, and they are both properly classed as
mini-BAL~QSOs. Although the C\,{\sc iv} absorption troughs are obviously broad 
($\Delta v\ga3000$\,km\,s$^{-1}$), these objects
do not formally meet the BAL~QSO criteria of Weymann et al. (1991).
With less extreme ultraviolet absorption, mini-BAL~QSOs might be
expected to be stronger X-ray sources than bona-fide BAL~QSOs.
In fact, the mini-BAL~QSO PG~1411+442 was successfully observed with
\emph{ASCA}; spectral analysis indicated a substantial intrinsic
absorber with a column density, $N_{\rm
H}\approx10^{23}$\,cm$^{-2}$, and an absorption covering fraction, $f_{\rm
cov}\approx97\%$ (Brinkmann et al. 1999; Gallagher et al. 2001).    

RX~J0911.4+0551 was discovered as part of a program to identify bright 
\emph{ROSAT} sources (Bade et al. 1997), though the 29.2\,ks \emph{Chandra} observation of 1999
Nov 2 showed it be a factor of
$\approx8$ fainter than during the \emph{ROSAT} All Sky Survey (Chartas et
al. 2001).  Spectral analysis revealed an X-ray continuum with a
typical photon index overlaid with absorption. However, the absorption 
was not adequately modeled with neutral gas.  The low-energy residuals from a
power-law fit above rest-frame 5\,keV suggested some 
complexity in the absorption such as would result from either a partially covering
 or an ionized absorber.  Both models were
significant improvements over the neutral absorber with the first
model being slightly preferred (Chartas et al. 2001).  The best-fitting intrinsic 
column density, $N_{\rm H}=2\times10^{23}$\,cm$^{-2}$, is the largest of
the four QSOs presented in this paper.

PG~1115+080 is notable for significant variability of the ultraviolet O\,{\sc vi} emission and
absorption lines (Michalitsianos, Oliversen, \& Nichols 1996) as well
as X-ray flux changes (Chartas 2000).
This target has been observed with \emph{Chandra} on two occasions, for 26.2\,ks 
on 2000 Jun 2 and 9.7\,ks on 2000 Nov 3.  The data were analyzed
following the procedure outlined above with a similar result as for RX~J0911.4+0551; a
partially covering absorber model with
$N_{\rm H}=4\times10^{22}$\,cm$^{-2}$ was preferred over neutral absorption.
PG~1115+080 has fairly good photon statistics (see Figure~3), thus making it a good target for
additional observations to investigate X-ray spectral variability.
%
\begin{figure}[t!]
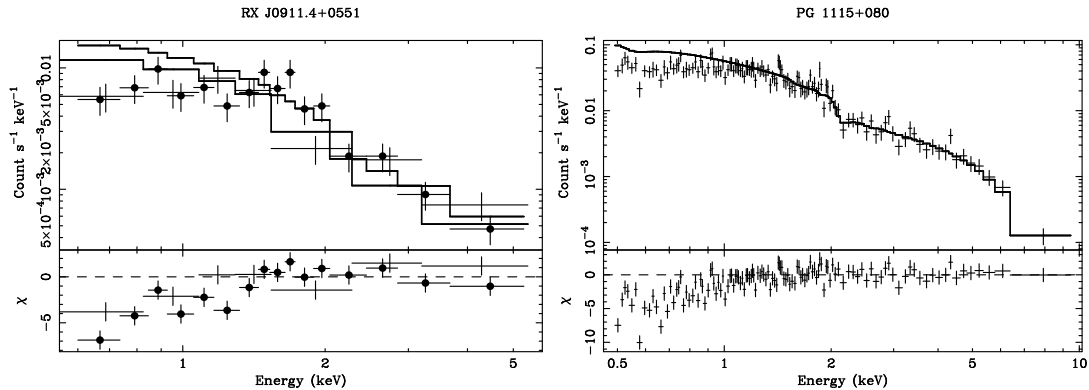

%
%
\centerline{\psfig{figure=gallagherfig3a.ps,height=2.0truein,width=2.8truein,angle=-90}
\psfig{figure=gallagherfig3b.ps,height=2.0truein,width=2.8truein,angle=-90}
}
{\caption{\protect \emph{Chandra} ACIS-S3 spectra of the two mini-BAL~QSOs, 
RX~J0911.4+0551 and PG~1115+080.  Both spectra have been fit above rest-frame 5\,keV
with a power-law model, which has then been extrapolated back to the lowest
energies.  A partial-covering absorption model provides a better fit
than neutral absorption for both objects.}}
\end{figure}
%
\section{General Picture and Conclusions}
As the number of BAL~QSOs detected with enough X-ray photons for spectral
analysis grows, a consistent picture is beginning to emerge.  The
X-ray continua can be well modeled by 
power-law models with photon indices consistent with those of other
radio-quiet QSOs, $\Gamma\approx2$.  In addition, correcting the X-ray spectra for
absorption reveals normal ultraviolet-to-X-ray flux ratios, thus indicating 
that the underlying spectral energy distributions of BAL~QSOs are not unusual.
Both of these observations support the scenario whereby broad
absorption line outflows are  common components of the nuclear
environments of radio-quiet QSOs.

Confirming these generalizations will require additional, long
spectroscopic observations of the brightest BAL~QSOs.  To complement
this endeavor, large, exploratory surveys of well-defined samples of
BAL~QSOs will offer enough information to examine which multi-wavelength
properties are related to the X-ray characteristics.  As of yet,
predicting which BAL~QSOs will be productive targets for spectroscopic 
X-ray observations remains a black art.
Connecting the X-ray properties, such as flux and coarse spectral shape, to
properties in other spectral regimes will ultimately help us to
understand the nature of the intrinsic absorption in the X-ray and
ultraviolet. We have begun such a program in the 
\emph{Chandra} Cycle~2 observing round with 18 BAL~QSO targets from the Large
Bright Quasar Survey, and the data that have arrived thus far are promising.

In terms of the column density of the absorbing gas, the best-fitting
values range from (1--20)$\times10^{22}$\,cm$^{-2}$.  Though relatively
simple models can adequately explain the observations, the physical
absorber in each system is likely to be complex.  Partial-covering absorption models
suggest that multiple lines of sight are present; the direct view
suffers from heavy obscuration while a second, scattered line of sight 
could be clearer.  In addition, if the X-ray absorbing gas is close to the
nucleus and associated with the ultraviolet-absorbing
gas, it must also be highly ionized. 
In this case, the column density measurements can only be considered
lower limits.  Additionally, significant velocity dispersion of the absorber would
increase the continuum opacity of bound-bound absorption lines, and thus further
complicate determining an accurate value of the column density.

X-rays are powerful probes of the inner regions of BAL~QSOs as they
are sensitive to molecular, neutral, and partially ionized gas.
Estimates of the absorption column density from X-ray observations
(compared to ultraviolet spectral analysis) 
suggest that the bulk of the absorbing gas is more readily accessible
in the high-energy regime.  Thus, X-ray 
observations offer the greatest potential for determining the true mass
outflow rate of QSOs.  To constrain this value, a measure of the
velocity structure of the X-ray absorbing gas is essential.  To this
end, a gratings observation with \emph{XMM-Newton} of the X-ray brightest BAL~QSO, 
PG~2112+059, offers the most promise.  

\acknowledgments
This research was supported by NASA grant NAS8-38252, Principal Investigator, GPG.
SCG also gratefully acknowledges NASA GSRP grant NGT5-50277.

\end{document}